# Idea Inheritance, Originality, and Collective Innovation

**Harris Kyriakou**
Stevens Institute of Technology
ckyriako@stevens.edu

**Jeffrey V. Nickerson**
Stevens Institute of Technology
jnickerson@stevens.edu

**INTRODUCTION**

In order to create new products, inventors *search* and *combine* previous ideas [10, 14, 17]. Few studies have examined the characteristics of search that lead to new products; most have focused on patent citations, which are often retrospective and may not reflect the usefulness of inventions [10].

Through the analysis of collaborations in an online virtual community, the impact of originality on popularity and practicality is tested. These tests in turn are based on a method for measuring the distance between 3D shapes. In sum, this paper presents a new method for gauging innovation, and suggests ways of further understanding the role technology plays in encouraging creativity. From an organization perspective, this work provides insights into the creative process, and in particular the open innovation process, in which thousands of individuals together evolve designs, without belonging to the same corporate structure, without claiming IP rights, without exchanging money.

**Background**

Thingiverse is a design social network in which participants can share their designs or combine (remix) preexisting designs. They can also show their interest in a design by liking it, or by printing it. If a user prints a design, the user will post a photo of the printed item back to the project. Thus, there are two ways design success can be measured: by the number of times it has been liked, a measure of popularity, and by the number of times it has been printed, called *makes*, a measure of practicality.

The objects themselves have certain qualities – some are similar to each other, and some are very different. One way to approximate the similarity is to measure the geometrical distance (the shape difference) between objects. Then, inventors are exploring a design space, a vast range of possible shapes. The space is too large for an individual designer to explore, but many independent designers may traverse the space in parallel. That is, they transform distant search – search that would involve huge leaps – into local search, because each individual can explore a region of the landscape [1].

This idea, which builds on older ideas of search [14], suggests an evolutionary theory of innovation: that children (new ideas) are born through a process of modification and recombination of parents (existing ideas). Children that are much different from all parents will attract more attention than children similar to their parents. That is, remixes distinct from original designs will attract more likes than imitative designs. But these original designs, untested as they are, will be less likely to be practical, and therefore will not be printed. Designs with no parent designs will also be less likely to be printed than those that inherit from other designs, as their features are less likely to have been tested than those that are embedded in the network.

Specifically, literature from many fields has found that people are attracted to novelty. In the context of Thingiverse, people are given the opportunity to show their approval by *liking* designs; these likes will indicate the attractiveness of the design. This leads to the following hypothesis:

*Novelty leads to popularity:* Original designs in a remix network will be more popular than imitative designs.

By a remix network we mean networks that allow search on a database of ideas. By original designs we mean designs that are far from existing designs, and by imitative designs we mean those close to preexisting designs. In our case we will measure the differences between the shapes of the design, as detailed later.

While novelty should drive popularity, literature on creativity suggests that many novel ideas are impractical [6, 2, 7]. And so we are led to the second hypothesis:

*Novelty leads to impracticality:* Novel designs in a remix network will be less practical than imitative designs, as measured by the number of times a design is instantiated.

The placement of a design in the overall remix network should also make a difference in its popularity and practicality. Centrality measures are important in other kinds of remix networks, such as open source communities [13]. The general consensus is that central designs are more likely to be successful, which in the open source community implies both popularity and practicality. In collective invention environments where the disclosure of information supports high innovation rates and fast knowledge accumulation, communication network structure has been found to have a strong influence on system performance [5]. Collective exploration improves average success over independent exploration because good solutions can propagate through the network [15].

Actors in brokerage positions in networks enjoy the benefits of creating value by bridging information and resources from distant groups [4]. In addition, it has been suggested that companies should focus on developing innovations that bridge different technologies as their technological developments reach maturity [18]. However, centrality is also often associated with power and influence [3, 12]. The betweenness centrality of a patent contributes to its success [9].

This leads to the third hypothesis:

*Embeddedness leads to popularity and practicality:* Designs with parent designs will be more popular (as measured by likes) and more practical (as measured by makes) than designs without parent designs.

## RESULTS

In an effort to apply a more objective way to measure originality we have used a method [11] of measuring the conceptual distance between designs primarily used in computer graphics literature. Hypotheses were tested on Thingiverse.

The distance measure was run on 16,139 designs from Thingiverse. The distances were split along the mean into a set of original and a set of imitative designs. A t-test was run comparing the mean number of likes in these two sets: The results are shown in the left column of Table 1, and on the left of Figure 1. The first hypothesis is supported: more original designs are more popular. A test was run comparing the mean number of makes in the original and imitative designs. The results are shown on the right column of Table 1, and the right of Figure 1. The second hypothesis is not supported: original designs are not less practical. Indeed, they are significantly more practical than imitative designs.

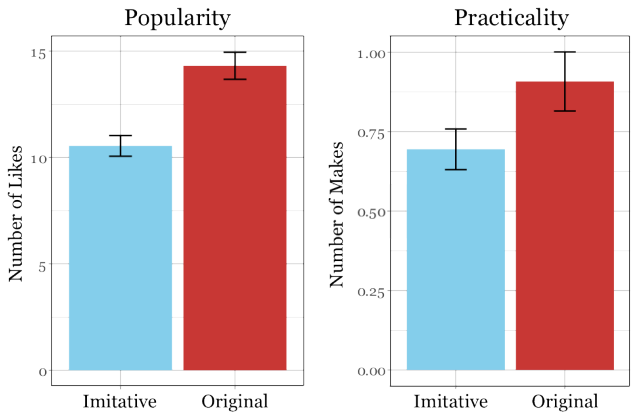

**Figure 1. Imitative vs. Original. Error bars show 95% confidence intervals.**

| **Originality** | Popularity | Practicality |
|---|---|---|
| t | 15.67 | 4.70 |
| df | 14,260.91 | 15,105.76 |
| p-value | < 2.2e-16 | 2.66e-06 |
| 95% C.I. | 4.86-6.25 | 0.13-0.33 |
| µ(Original) | 15.80 | 0.94 |
| µ (Imitative) | 10.24 | 0.71 |

**Table 1. Originality -Welch Two Sample t-test**

To test the third hypothesis, which suggested that embeddedness leads to popularity and practicality, we split all designs into two categories, designs with no parents (Standalone designs), and designs with one or more parents (Inherited designs). A t-test was performed comparing mean likes, and another comparing mean makes. The results are shown in Figure 2 and Table 2. The hypothesis is supported.

## Discussion and Conclusions

We have utilized a method for measuring originality by computing the distance between products, as a function of shape distances. Using this measure, we showed that, consistent with a general evolutionary theory of product design, original designs were more popular, as measured by the number of likes that they received. But we also found that original designs were more practical, as measured by the number of designs that were actually printed. This runs counter to the perceived tradeoff between exploration and exploitation. At least in the 3D printing world, explorers are rewarded more than exploiters, meaning their work is both liked more and printed more.

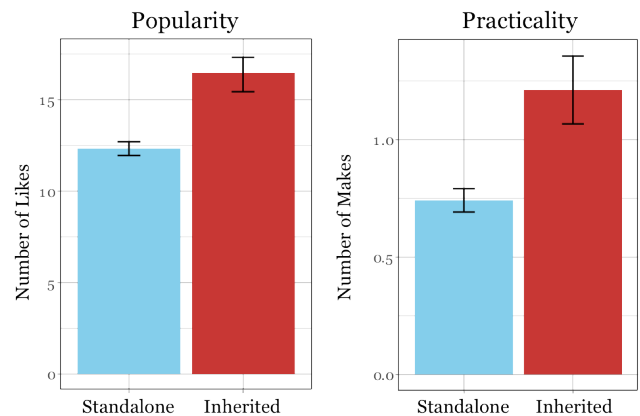

**Figure 2. Standalone vs. Inherited. Error bars show 95% confidence intervals**

| **Inheritance** | Popularity | Practicality |
|---|---|---|
| t | 8.01 | 6.03 |
| Df | 3874.08 | 3610.33 |
| p-value | 1.492e-15 | 1.826e-09 |
| 95% C.I. | 3.13-5.16 | 0.32-0.62 |
| µ(Inherited) | 16.47 | 1.21 |
| µ(Standalone) | 12.33 | 0.74 |

**Table 2. Inheritance - Welch Two Sample t-test**

Why might this be? It could be that the decision to print something is a function of many factors, and one of them may be related to originality. That is, if one already has an object, then one is less likely to need a duplicate. Thus, more original objects are more likely to be printed. But it could also be that better designers consider both practicality and originality when they work, and therefore are unlikely to produce imitative work. Understanding this phenomenon will require more study, and possibly experimentation, in order to understand what drives users to print designs, as well as what drives inventors to remix specific designs.

A third finding is that designs that inherit from others are both more popular and more practical. This finding provides support for our intuitions that sharing information in networks (for example, the way academics share papers) leads to better results. It also raises further questions: do designers increase their inheritance behavior as they engage further with the community? Are solo designers more likely to drop out of the community? These questions are particularly relevant to online community designers: remix communities provide certain affordances, and it will be useful to know which affordances lead to increased participation, as well as the evolution of better products.

It has been suggested that it is easier for an organization to simultaneously excel at exploration and exploitation than it would be for an individual [8]. It will be useful to know how much better, and whether the loose organizations called online communities are better or worse environments than the R&D departments of companies. If innovation networks thrive as organizational forms when the sources of industry expertise are widely dispersed and the knowledge base is complex and expanding [16], then we expect in more complex environments that the positive effects of network embedding will be even greater and that open innovation networks may provide a unique opportunity for rapid creation of complex systems at a minimal cost.

The way forward may take multiple paths. There is more to be done in understanding how remix networks grow, and what contributes to the behavior of the members of these networks. Open innovation usually draws a distinction between designers and users who offer design advice: in the 3D community discussed here these boundaries are very much blurred, as members can consume, make, and modify designs as they wish. There is a notion of co-design, but it not a co-design between a designer and several consumers; it is a co-design in which thousands of designers are interacting and improving on each other's work. From an information systems perspective, these networks are useful to study: they provide traces of organizational behavior that often remain invisible (or don't exist) inside companies. The collaborative technologies are definitely affecting behavior. The systems affordances – the facilities that encourage design – can be studied in relationship to the traces of behavior left behind. So can the product results that emerge from the behavior, individual and collaborative.

There are challenges: the openness of these systems, the permeability, means the phenomena will be changing over time, and so statistics about these environments will likely shift over time. On the other hand, it may be possible to learn what kinds of environments encourage faster and better evolution of products, and, at a higher level, what affordances advance the evolution of the community.

**ACKNOWLEDGEMENTS**
This material is based upon work supported by the National Science Foundation under Grant Number 0968561.